\documentclass{spie}
\usepackage{graphicx}
\graphicspath{{images/}}
\usepackage[euler]{textgreek}

\title{Volume phase holographic gratings for the Subaru Prime Focus Spectrograph: performance
measurements of the prototype grating set}

\author{Robert Barkhouser\supit{a}\textsuperscript{*}, James Arns\supit{b} and James E. Gunn\supit{c}
\skiplinehalf
\supit{a}Johns Hopkins University, Department of Physics \& Astronomy, 3400 N Charles Street, Baltimore, MD 21218, USA; \\
\supit{b}Kaiser Optical Systems, Inc., 371 Parkland Plaza, Ann Arbor, MI 48103, USA; \\
\supit{c}Princeton University, Department of Astrophysical Sciences, 4 Ivy Lane, Princeton, NJ 08544, USA}

\authorinfo{\textsuperscript{*}Email:  rhb@jhu.edu, Telephone:  1 410 516 7918}

\begin{document}
\maketitle

%%%%%%%%%%%%%%%%%%%%%%%%%%%%%%%%%%%%%%%%%%%%%%%%%%%%%%%%%%%%%
\begin{abstract}
The Prime Focus Spectrograph (PFS) is a major instrument under development for the 8.2 m Subaru telescope on Mauna Kea. Four identical, fixed spectrograph modules are located in a room above one Nasmyth focus. A 55~m fiber optic cable feeds light into the spectrographs from a robotic fiber positioner mounted at the telescope prime focus, behind the wide-field corrector developed for HyperSuprime Cam. The positioner contains 2400 fibers and covers a 1.3~degree hexagonal field of view. Each spectrograph module will be capable of simultaneously acquiring 600 spectra.

The spectrograph optical design consists of a Schmidt collimator, two dichroic beamsplitters to separate the light into three channels, and for each channel a volume phase holographic (VPH) grating and a dual-corrector, modified Schmidt reimaging camera. This design provides a 275~mm collimated beam diameter, wide simultaneous wavelength coverage from 380~nm to 1.26~\textmu m, and good imaging performance at the fast f/1.05 focal ratio required from the cameras to avoid oversampling the fibers. The three channels are designated as the blue, red, and near-infrared (NIR), and cover the bandpasses 380--650~nm (blue), 630--970~nm (red), and 0.94--1.26~\textmu m (NIR). A mosaic of two Hamamatsu 2k$\times$4k, 15~\textmu m pixel CCDs records the spectra in the blue and red channels, while the NIR channel employs a 4k$\times$4k, substrate-removed HAWAII-4RG array from Teledyne, with 15~\textmu m pixels and a 1.7~\textmu m wavelength cutoff.

VPH gratings have become the dispersing element of choice for moderate-resolution astronomical spectrographs due their potential for very high diffraction efficiency, low scattered light, and the more compact instrument designs offered by transmissive dispersers.  High quality VPH gratings are now routinely being produced in the sizes required for instruments on large telescopes.  These factors made VPH gratings an obvious choice for PFS.  In order to reduce risk to the project, as well as fully exploit the performance potential of this technology, a set of three prototype VPH gratings (one each of the blue, red, and NIR designs) was ordered and has been recently delivered. The goal for these prototype units, but not a requirement, was to meet the specifications for the final gratings in order to serve as spares and also as early demonstration and integration articles.  In this paper we present the design and specifications for the PFS gratings, the plan and setups used for testing both the prototype and final gratings, and results from recent optical testing of the prototype grating set.
\end{abstract}

%>>>> Include a list of keywords after the abstract


%%%%%%%%%%%%%%%%%%%%%%%%%%%%%%%%%%%%%%%%%%%%%%%%%%%%%%%%%%%%%
\section{INTRODUCTION}
\label{sec:intro}  % \label{} allows reference to this section

Volume phase holographic (VPH) gratings have been under development for decades\cite{Arns95} and their potential for use in astronomical spectrographs was put forth by Barden et~al. in 1998\cite{Barden98}.  Larger instruments for larger telescopes have fueled the demand for large, high quality VPH gratings but until recently their availability has been very limited.  Difficulties have included the ability to coat and process high quality holographic emulsions on large glass plates, the size and quality of the optics required to create the holographic exposure system, and the ability to keep such a large interferometric system stable enough to produce good exposures.  However, film coating and processing quality has steadily improved, and Kaiser has invested the time and equipment required to reliably produce high quality gratings up to 300~mm in diameter in a single (i.e., non-mosaiced) exposure.

For PFS, a total of 16 gratings will be required.  Four identical spectrograph modules each contain three spectroscopic channels designated as the blue (380--650~nm), red (630--970~nm), and near-infrared (NIR, 0.94--1.26~\textmu m).  Each channel employs a single grating which disperses the full bandpass onto the detectors, and the red channel will contain an additional, medium resolution grating that is part of a grism assembly and resides in an exchange mechanism along with the low resolution grating.  The medium resolution gratings will be the last to be fabricated and are not discussed in this paper.

A set of three VPH gratings (one each of the blue, red, and NIR designs) has been successfully fabricated and delivered. These first units are considered prototypes, and the goal was to meet the specifications for the actual gratings in order to serve as spares and also as early demonstration and integration articles.  In this respect, the prototype gratings have been extremely useful.  While not fully compliant with all the specifications for the production gratings, they do come close and have provided very important insights into both the manufacturing process and our testing program.  We believe the insights gained from these prototype units will allow PFS to deploy the highest quality, large-format VPH gratings produced to date.

%%%%%%%%%%%%%%%%%%%%%%%%%%%%%%%%%%%%%%%%%%%%%%%%%%%%%%%%%%%%%
\section{PFS PROJECT OVERVIEW}

PFS is a multi-object, 380~nm to 1.26~\textmu m spectrograph being developed for the 8.2~m Subaru telescope\cite{Sugai12}.  PFS is part of the Subaru Measurement of Images and Redshifts (SuMIRe) project, providing the spectroscopic capability while while Hyper Suprime-Cam (HSC) provides the imaging capability. PFS and HSC share the Wide Field Corrector (WFC), and the excellent image quality delivered by HSC has demonstrated the high quality of the WFC.

A robotic fiber positioner with 2400 fibers will be located at prime focus behind the WFC, covering a 1.3~degree, hexagonally-shaped field.  A long fiber optic cable carries light from the prime focus down to four identical spectrograph modules located in an insulated, cooled room on the IR4 floor of the Subaru dome.  The spectrographs are bench-mounted and the temperature of the room is controlled to $5\pm1\,^{\circ}{\rm C}$.  This allows the use of near-room-temperature spectrograhs while maintaining a sufficiently low thermal background to operate out to 1.26~\textmu m.

Each spectrograph module receives 600 fibers and produces a 275~mm diameter beam using an f/2.5 Schmidt collimator.  The beam is split via dichroic beamsplitters into the three channels: blue, red, and NIR.  The full 380~nm to 1.26~\textmu m bandpass is dispersed with an average resolving power of 3000 by the VPH gratings and recorded simultaneously onto 15~\textmu m pixel CCD and HgCdTe detectors.  The blue and red channels will use a mosaic of two Hamamatsu 2k$\times$4k, fully-depleted CCDs, while the NIR channel will use a 4k$\times$4k, substrate-removed H4RG array from Teledyne.  In order to reasonably sample the fibers with 15~\textmu m pixels, very fast cameras are required and PFS will employ f/1.09, 300~mm focal length Schmidt cameras.  The spectrograph optical design is described in another paper in these proceedings\cite{Pascal14}.

%%%%%%%%%%%%%%%%%%%%%%%%%%%%%%%%%%%%%%%%%%%%%%%%%%%%%%%%%%%%%
\section{GRATING DESIGN}
\label{sec:design}

The PFS gratings are 340$\times$340~mm in outline.  They are made from two 20~mm thick BK7 plates, bonded together to form a 40~mm thick sandwich with the grating layer buried in the middle.  One plate, the substrate, contains the dichromated gelatin layer into which the grating is holographically recorded. The cover plate is bonded to the substrate after the grating layer has been exposed, processed, and verified.  Unique to PFS is the inclusion of a pupil stop as part of the grating assembly itself, consisting of a thin layer of patterned chrome on the interior surface of the cover plate.  The following sections describe the grating design in more detail.

%%-----------------------------------------------------------
\subsection{Grating parameters}
\label{sec:parameters}

\begin{figure}[]
    \centering
    \includegraphics[height=11cm]{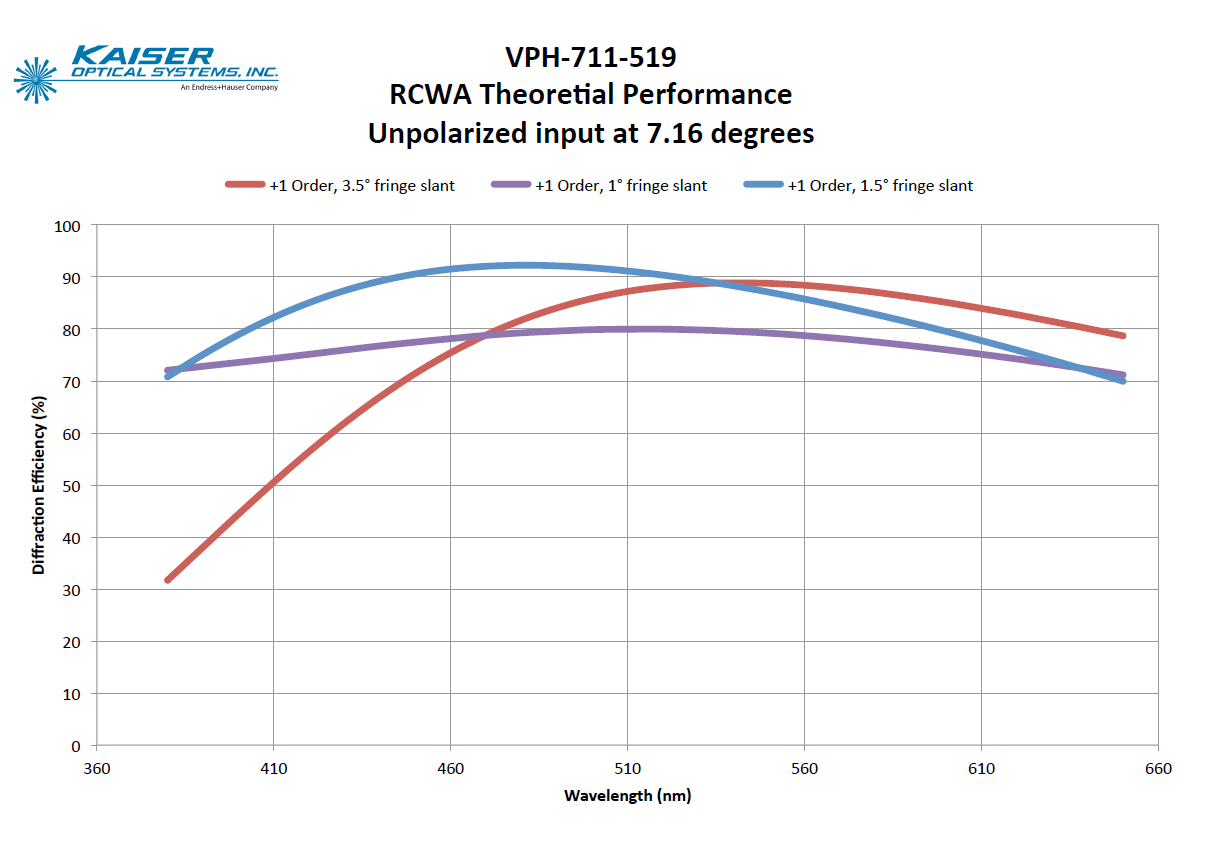}
    \caption{Theoretical diffraction efficiency prediction for the blue grating.}
    \label{fig:blue_rcwa}
\end{figure}

\begin{figure}[]
    \centering
    \includegraphics[height=11cm]{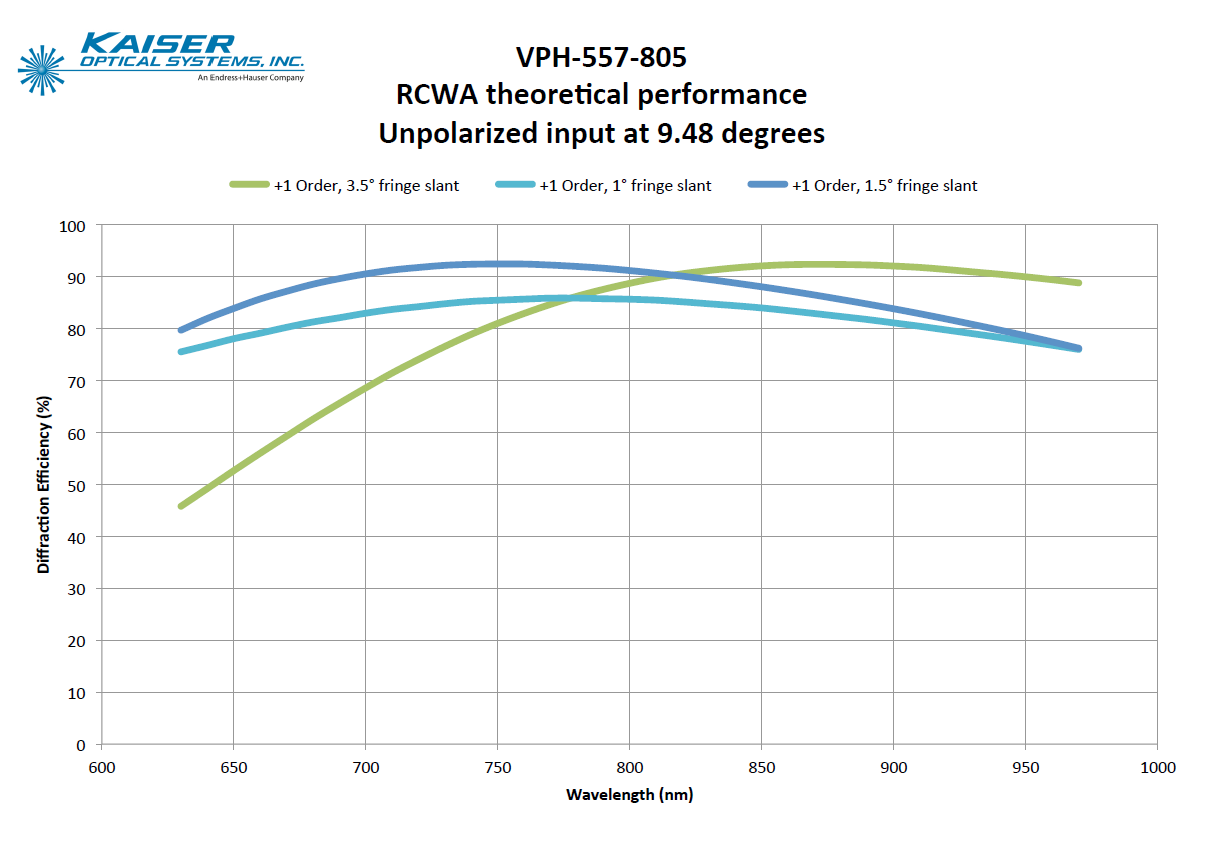}
    \caption{Theoretical diffraction efficiency prediction for the red grating.}
    \label{fig:red_rcwa}
\end{figure}

\begin{figure}[]
    \centering
    \includegraphics[height=11cm]{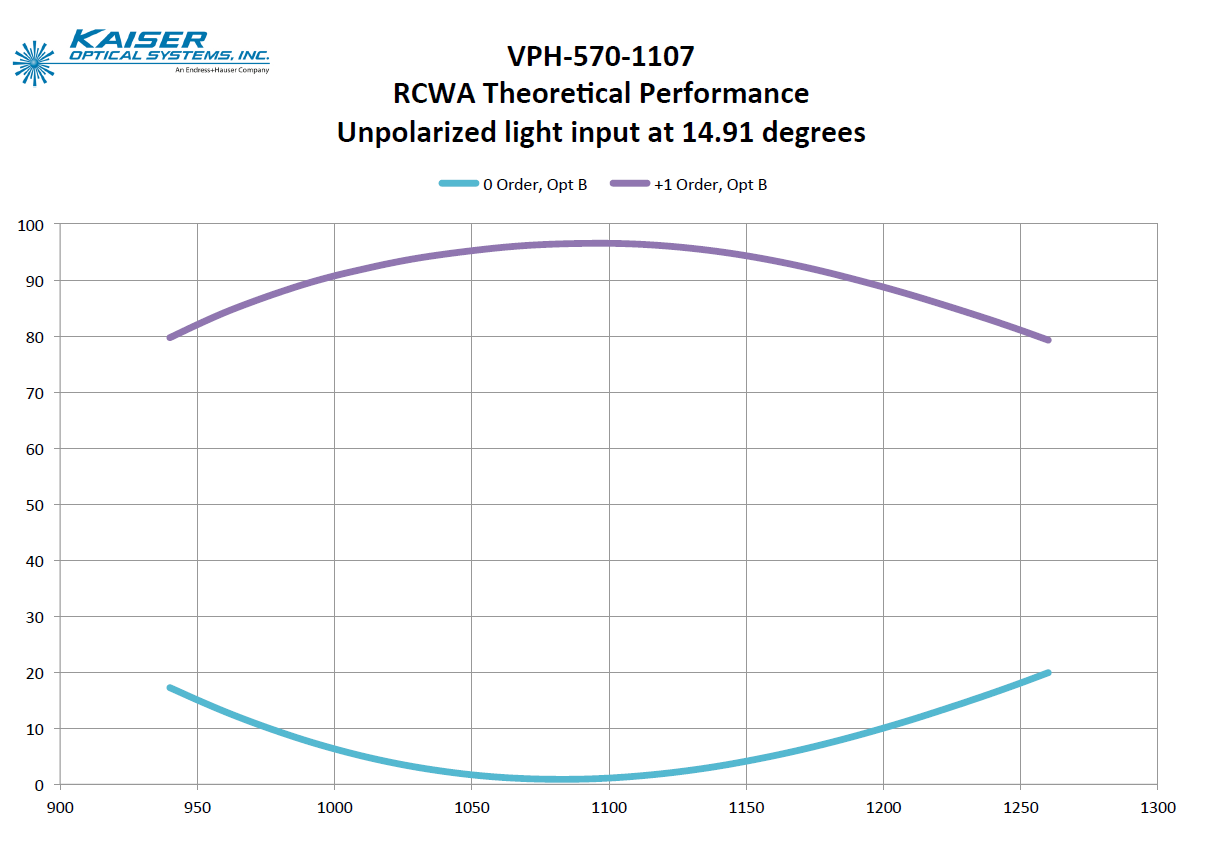}
    \caption{Theoreticial diffraction efficiency prediction for the NIR grating.}
    \label{fig:nir_rcwa}
\end{figure}

The basic grating parameters are given in Table \ref{tab:parameters}.  They are derived from the spectrograph optical design, which of course must deliver the resolving power and wavelength coverage necessary to meet the science requirements.  In addition, the gratings will be operated in an off-Littrow configuration to avoid the well-known Littrow ghosts (see \S \ref{sec:ghost} below).  Given these basic design parameters, rigorous coupled wave analysis (RCWA) was used to explore the grating design space available with input parameters that are available and reasonable in the production of VPH gratings.  Theoretical efficiency curves resulting from RCWA for each grating design (blue, red, and NIR) are presented in Figures~\ref{fig:blue_rcwa}, \ref{fig:red_rcwa}, and \ref{fig:nir_rcwa}.

\begin{table}[]
    \centering
    \caption{PFS grating design parameters.}
    \medskip
    \begin{tabular}{lccc}
        \textbf{Parameter} & \textbf{Blue} & \textbf{Red} & \textbf{Near-IR} \\
        \hline
        Substrate material & N--BK7 & N--BK7 & N--BK7\\
        Substrate dimensions & 340$\times$340 mm & 340$\times$340 mm & 340$\times$340 mm\\
        Overall thickness & 40 mm & 40 mm & 40 mm\\
        Minimum clear aperture & 275 mm & 275 mm & 275 mm\\
        Wavelength range & 380--650 nm & 630--970 nm & 940--1260 nm\\
        Central wavelength & 519 nm & 805 nm & 1107 nm\\
        Fringe frequency & 711.4 mm$^{-1}$ & 556.7 mm$^{-1}$ & 569.5 mm$^{-1}$\\
        Fringe slant & ~1.5 deg & ~1.0 deg & ~2.0 deg\\
        Angle of incidence (\textalpha) & ~7.2 deg & ~9.5 deg & 14.9 deg\\
        Angle of diffraction (\textbeta) & 14.2 deg & 16.5 deg & 21.9 deg\\
    \end{tabular}
    \label{tab:parameters}
\end{table}

%%-----------------------------------------------------------
\subsection{Pupil masks}

One unique feature of the PFS gratings is the inclusion of a pupil mask as an inherent part of the grating assembly itself.  The optical design of the spectrograph provides a single pupil location, which is at the grating.  However, the grating is buried 20~mm deep within the substrate/cover plate sandwich.  The collimated beam diameter is already at the limit of what Kaiser is able to expose without straying outside the optically specified clear apertures of the exposure optics.  So it is important that the pupil be located at the grating layer within the glass.  However, this means the pupil is not accessible for placement of a system stop, and any stop external to the grating would need to be undersized in order to keep light within the clear aperture of the grating, and, as well, the other optics in the system (which clear apertures have been specified based on the clear aperture of the grating).  Otherwise there is the risk that light falling outside the specified clear aperture of any optic ends up in the wings of the PSF due to excessive wavefront or surface error.

\begin{figure}[h]
    \centering
    \includegraphics[height=10cm]{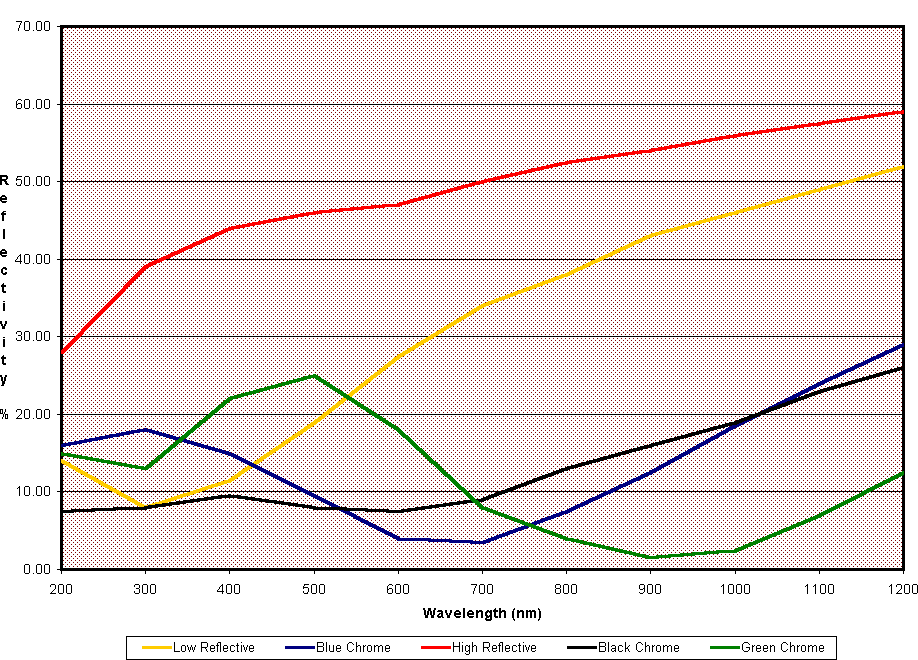}
    \medskip
    \caption{Reflectivity curves for various chrome coatings.}
    \label{fig:chrome_refl}
\end{figure}

In an effort to solve this problem, we investigated the feasibility of placing a low reflectivity chrome mask on the interior surface of the cover plate which gets bonded to the grating substrate.  Applied Image\footnote{Applied Image, 1653 East Main Street, Rochester, NY 14609 USA} fabricated several reduced scale masks on 125~mm square BK7 glass plates, and the results were very good.  Kaiser then bonded these samples to other plates and put them through environmental testing to insure the integrity of the bond joint in the presence of the chrome.  Figure~\ref{fig:chrome_refl} shows the reflectivity curves for several different types of chrome coatings.  However, at this time the only type we have been able to source is the blue chrome, which is perfect for the red channel and acceptable for the blue and NIR.

\begin{figure}[b]
    \centering
    \includegraphics[height=7cm]{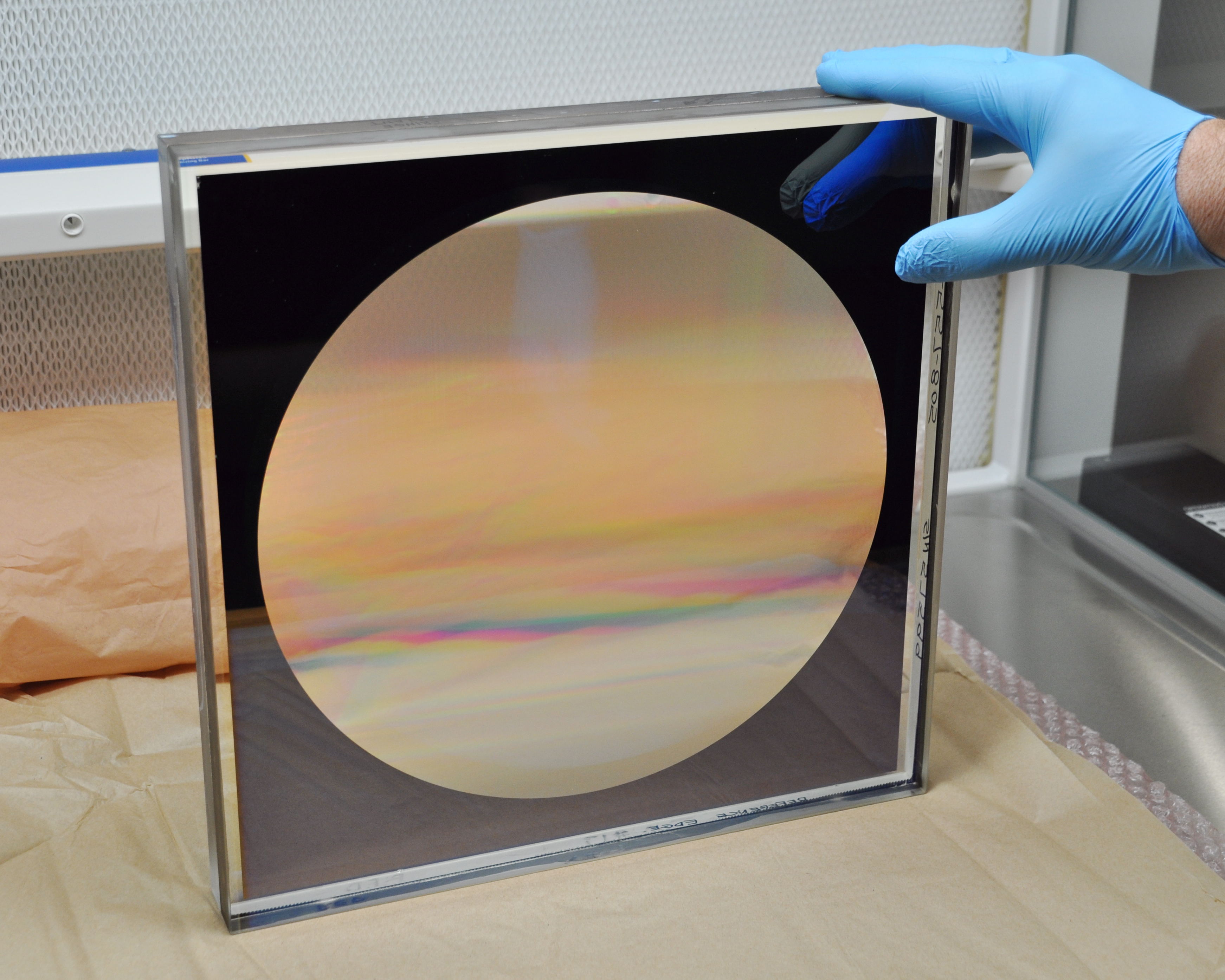}
    \caption{Photograph of red VPH grating.}
    \label{fig:red_photo}
\end{figure}

The full-sized masks are slightly elliptical due to the tilt of the gratings with respect to the beam.  We chose to use a common mask size for all the gratings, 275 x 282~mm, because the maximum difference in incidence angle is only 7.7 degrees and the percentage of aperture optimized with unique masks is very small.  The chrome layer does not extend all the way to the edge of the glass; there is a 10~mm border of clear glass on all sides to ensure a good bond joint and seal at the edges. The large masks have been more problematic to fabricate, however, in part due to their size and also the presence of the AR coating on one side of the plate, which requires protection during processing and also limits the available cleaning methods used prior to deposition of the chrome layer and necessary to achieve a good chrome layer.  The masks produced for the prototype gratings did have a large number of pinholes and were not completely opaque.  However, we estimated the transmission through the masks to be less than 1\% and that is sufficient for our purposes.  Figure~\ref{fig:red_photo} is a photograph of the completed red prototype grating showing the chrome mask.

%%-----------------------------------------------------------
\subsection{Littrow ghost and fringe tilt}
\label{sec:ghost}

Any spectrograph with a grating used in Littrow mode (\textalpha~=~\textbeta) is susceptible to the so-called ``Littrow ghost'' which Burgh et al. termed ``recombination ghost'' in their paper which shed much light on the origins of, and remedies for, this troublesome apparition\cite{Burgh07}.  It is a white light ghost originating from a reflection of dispersed light off the detector, which then travels backwards through the camera.  The camera recollimates the light and the grating recombines the collimated light, producing a white light ghost beam traveling backward out of the grating.  From this point a planar surface (e.g. the grating entrance face) can provide the second ghost bounce and send a portion of the beam back through the grating in zeroth order, undispersed.  The camera will then focus the ghost back onto the detector, at the location of the Littrow wavelength.

For VPH gratings, the diffraction efficiency is governed by the Bragg condition, and for a volume phase grating whose fringe planes are normal to the grating surface, the Bragg condition is satisfied with the grating in Littrow configuration where $m\nu\lambda=2sin(\alpha)$.  But this configuration leads to a recombination ghost right at the center of the detector, assuming the Littrow wavelength is centered on the detector. One approach to mitigate the recombination ghost is to tilt the grating off of Littrow, enough to move the Littrow wavelength off the detector entirely or else to a location on the detector which receives no useful photons.  Off Littrow, however, the Bragg condition will no longer be satisfied and the diffraction efficiency will suffer.  In order to recover the Bragg condition, the fringe planes within the grating must be tilted to compensate for the off-Littrow tilt of the grating.  This is commonly referred to as ``fringe slant'' and is a technique which has been employed by Kaiser for many years.

For PFS the gratings will all be tilted 3.5 degrees off Littrow to move the ghost off the detectors.  The RCWA charts in Figures~\ref{fig:blue_rcwa} and \ref{fig:red_rcwa} show theoretical efficiency curves for various fringe slant angles.  Clearly, slanting the fringes by the same amount as the grating tilt is not always the best thing to do.

%%%%%%%%%%%%%%%%%%%%%%%%%%%%%%%%%%%%%%%%%%%%%%%%%%%%%%%%%%%%%
\section{TEST PLAN AND MEASUREMENT SYSTEMS}

Optical testing of the PFS gratings takes place both at Kaiser and at Johns Hopkins University (JHU).  Measurements at Kaiser include diffraction efficiency and wavefront error; measurements at JHU include diffraction efficiency, wavefront error, and point spread function (PSF).  The testing done at Kaiser is directed primarily toward in-process assessment and final acceptance testing, while the JHU tests are designed to fully characterize the gratings for detailed analysis of the impact on overall instrument performance.  A very important secondary role of the prototype gratings has been to provide early test articles to ``shake out'' the setups and test approaches used at JHU.

%%-----------------------------------------------------------
\subsection{Diffraction efficiency}

Diffraction efficiency measurements are made using two different methods at Kaiser, and will be made using a full aperture measurement system at JHU.  Following are the details of these different methods.

\subsubsection{Diffraction efficiency at Kaiser using a white light source and spectrometer}
\label{sec:spectrometer}

\begin{figure}[h]
    \centering
    \includegraphics[height=9cm]{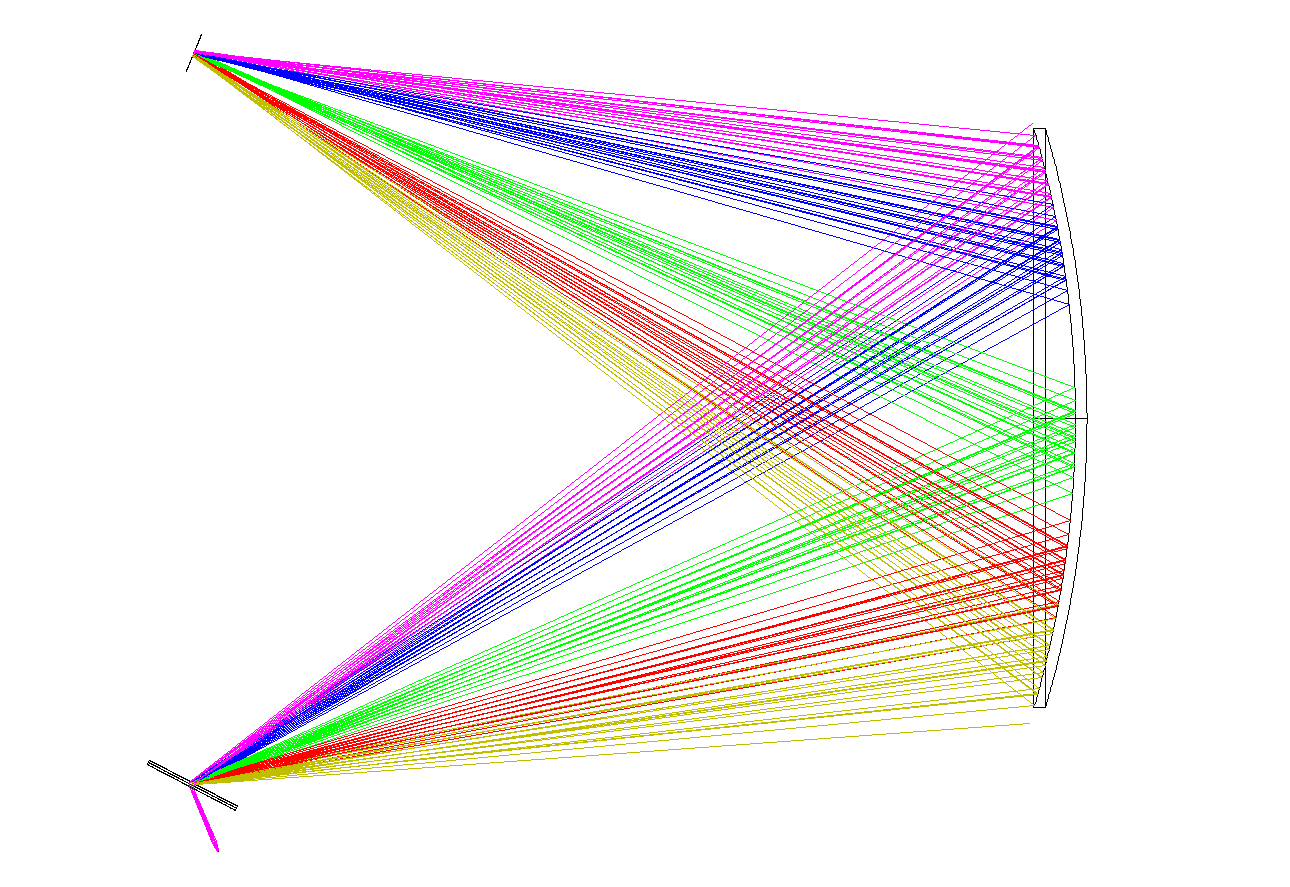}
    \caption{Schematic optical layout of the spectrometer based grating test method. The grating under test is located at lower left. The spherical mirror (right) and the entry port to an integrating sphere (upper left) form the basic collection optics.}
    \label{fig:de_kosi_1}
\end{figure}

The first method at Kaiser utilizes a white light source in the form of a deuterium/tungsten halogen lamp, a large collection mirror, an integrating sphere, and a spectrometer; see Figure~\ref{fig:de_kosi_1}.  The lamp and spectrometer are from Ocean Optics, a DH-2000 and HR4000 respectively. The primary collection optic is a 20~inch diameter, f/2 spherical mirror, and the integrating sphere is from LabSphere. A reference measurement of the light source and collection optics is taken without the grating in place. The grating is mounted on a translation/rotation stage that positions the grating to permit illumination at a variety of input angles and locations. After dividing the reference data from the raw sample data the diffraction efficiency of the grating is determined. Performance is measured at several physical locations to find the spatial uniformity. It is also possible to collect performance data at different input angles to find the angular bandwidth of the grating.

There are limitations of testing a VPH grating using this method. In particular, at the upper and lower wavelength extremes the spectrometer is starved for photons and the signal-to-noise is reduced to the point that the data is unusable. Measurements with this system are generally valid between 400 and 950~nm depending on the diffraction efficiency and spectral bandwidth of the grating under test.

\subsubsection{Diffraction efficiency at Kaiser using discrete lasers lines and ratiometer}
\label{sec:ratiometer}

The second method of measurement at Kaiser uses lasers with outputs at discrete wavelengths. The figure below is a schematic of the system. The light is depolarized and passed through an optical chopper. The light is split into two paths, one directed to a reference detector and the second to the VPH grating location.  The beam intensity is measured without the grating installed, representing the light incident on the grating. The grating is then mounted on a rotation stage and set to retro-reflect the interrogation light, establishing the reference angle. The grating is then rotated to the design reconstruction angle, in this case 20 degrees, and a detector is placed to intercept the +1 order diffracted light. Both detectors are input to a calibrated ratiometer that removes from the displayed value any variation in the laser intensity. The diffraction efficiency is calculated by dividing the reading from the +1 order diffracted light by the reading from the incident light.

\begin{figure}[]
    \centering
    \includegraphics[height=9cm]{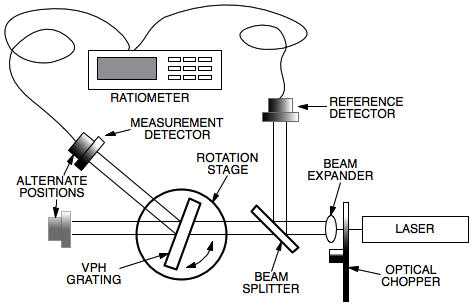}
    \caption{Schematic of diffraction efficiency measurement setup using a discrete wavelength laser.}
    \label{fig:de_kosi_2}
\end{figure}

\subsubsection{Diffraction efficiency measurements at JHU}

Measurements of VPH grating diffraction efficiency have been carried out at JHU previously, including gratings and assembled grisms for LDSS-3, and the upgraded SDSS spectrographs for BOSS\cite{smee13}.  These measurements were made with a small reflectometer consisting of a monochromator, collimating lens producing a 1~cm beam, reimaging lens, and a large-area (1~cm diameter) silicon photodiode.  Between the collimating and reimaging lenses was the grating holder, mounted on a rotation stage attached to a linear slide, allowing the grating to be set to the desired incidence angle and to be moved in and out of the beam.  The reimaging lens and photodiode were mounted on a common rail which pivoted about the axis of the rotation stage.  In this way a direct measurement of the monochromator flux could be made with the grating out of the beam and the detector arm positioned in line with the collimated beam.  Then the grating could be moved into the beam, the detector arm pivoted to the proper diffracted angle, and the photodiode reading compared to the direct measurement.  As long as the monochromator light source is stable during the time between direct and diffracted readings, and care is taken to control stray light in the setup, an accurate measurement (\textless~2\%) of the grating efficiency can be obtained.

This is the setup we planned to implement at JHU to measure the PFS gratings.  An obvious drawback of this system is that it only considers a tiny fraction of the grating aperture during any single measurement.  To sample many locations across the aperture would require many measurement runs (time consuming), and vertical adjustment of either grating mount or the beam height (cumbersome to implement with such large gratings).  Once we devised the setup for testing PFS, we realized we could use the same setup to easily make a full-aperture efficiency measurement in one shot.  This setup is described in detail below in \S \ref{sec:psf}.  We plan to measure each grating in steps of 10~nm over the full bandpass, in both zeroth and first orders.

%%-----------------------------------------------------------
\subsection{Wavefront error}

At Kaiser, wavefront measurements are made with 100~mm Zygo interferometer.  For large gratings such as those for PFS, this is a limiting factor as errors with spatial periods on the order of a few cycles per diameter (such as astigmatism, coma, etc.) can be completely missed.  However, wavefront errors with smaller spatial scales can be caught before the grating fabrication has been proceeded beyond the point of no return (i.e., capping).

\begin{figure}[]
    \centering
    \includegraphics[width=\textwidth]{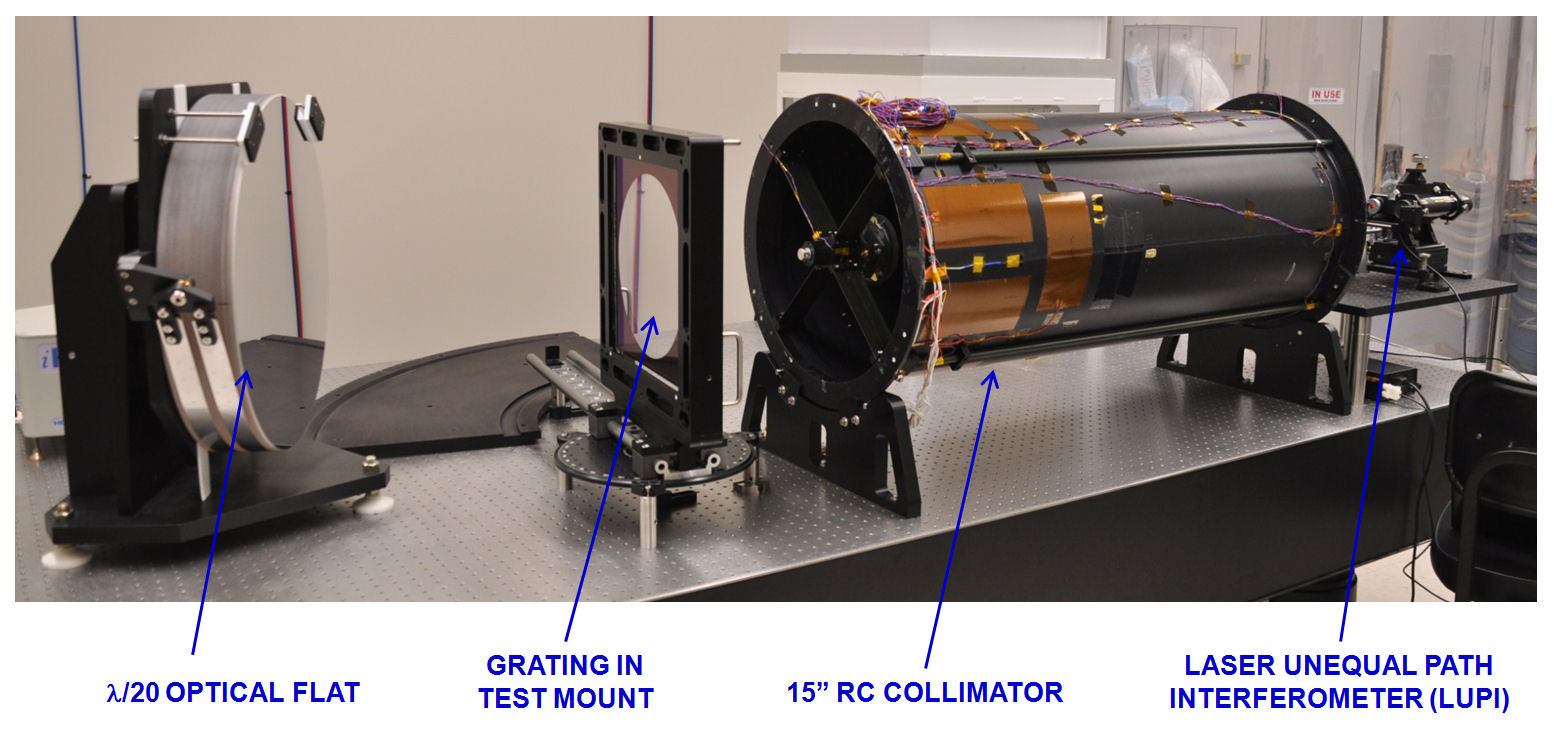}
    \caption{Setup for measuring transmitted wavefront error.}
    \label{fig:wfe_setup}
\end{figure}

The setup for measuring wavefront error at JHU consists of an interferometer with a diverger optic to bring the beam to focus at the focal plane of a collimator.  The expanded beam out of the collimator is incident on the grating at the desired angle, and the diffracted (or non-diffracted zeroth-order) beam is then autoreflected by a high quality optical flat.  This provides a classic double-pass setup, returning the beam back through the grating and into the collimator, which focuses the beam and sends it back into the interferometer for comparison against the internal reference beam.  Figure~\ref{fig:wfe_setup} is a photograph showing the major components of the wavefront measurement system.

\begin{figure}[]
    \centering
    \includegraphics[height=9cm]{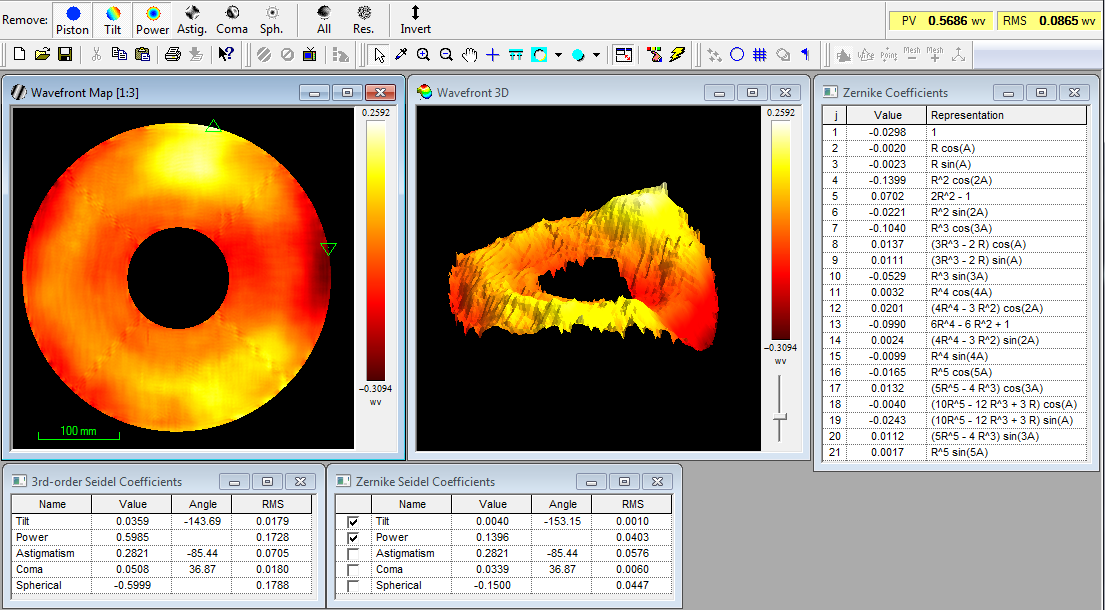}
    \caption{Wavefront error measured for the 15~inch RC collimator.}
    \medskip
    \label{fig:coll_wfe_full}
\end{figure}

\begin{figure}[]
    \centering
    \includegraphics[height=9cm]{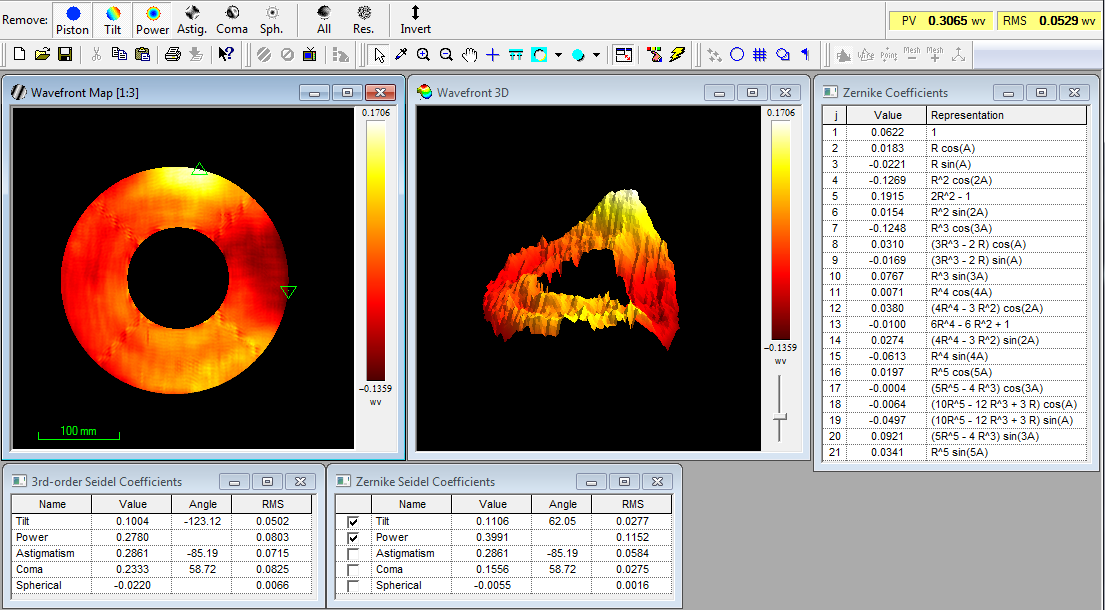}
    \caption{Wavefront error measured for the 15~inch RC collimator, stopped down to the 280 mm aperture of the gratings.}
    \label{fig:coll_wfe}
\end{figure}

The interferometer is a Buccini\footnote{Buccini Instrument Company, 118 Braxlo Lane, Wilmington, NC 28409 USA} MIC-1 laser unequal path interferometer\cite{Houston67} (LUPI), employing a modified Twyman-Green configuration with a 1~cm diameter internal beam and an f/10 diverging lens to illuminate the collimator.  The collimator is a two mirror, f/12 Ritchey-Chretien design with a 15~inch aperture and a 5~inch central obscuration.  It was designed at JHU in the mid-1990's for thermal-vacuum testing of a space-based instrument, and is capable of near diffraction-limited performance when carefully aligned.  Figure~\ref{fig:coll_wfe_full} shows the measured wavefront error of the collimator at full aperture, about 0.57~waves P-V and 0.09 waves RMS (\textlambda~=~632.8~nm).  Stopped down to the 280~mm clear aperture of the PFS gratings, Figure~\ref{fig:coll_wfe} shows the wavefront error to be 0.31~waves P-V and 0.05 waves RMS (\textlambda~=~632.8~nm).

\begin{figure}[]
    \centering
    \begin{tabular}{cc}
        \includegraphics[width=7cm]{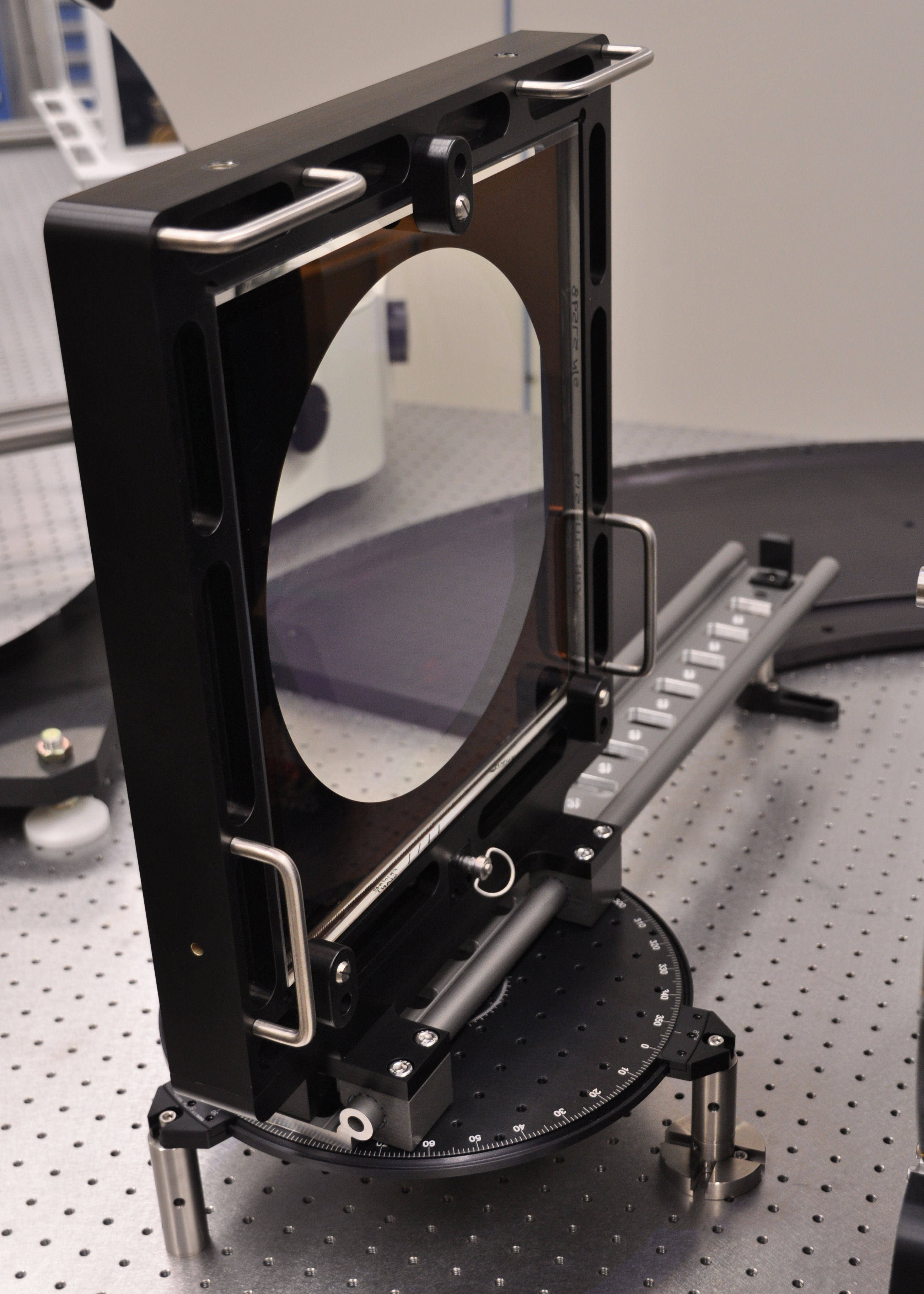} &
        \includegraphics[width=7cm]{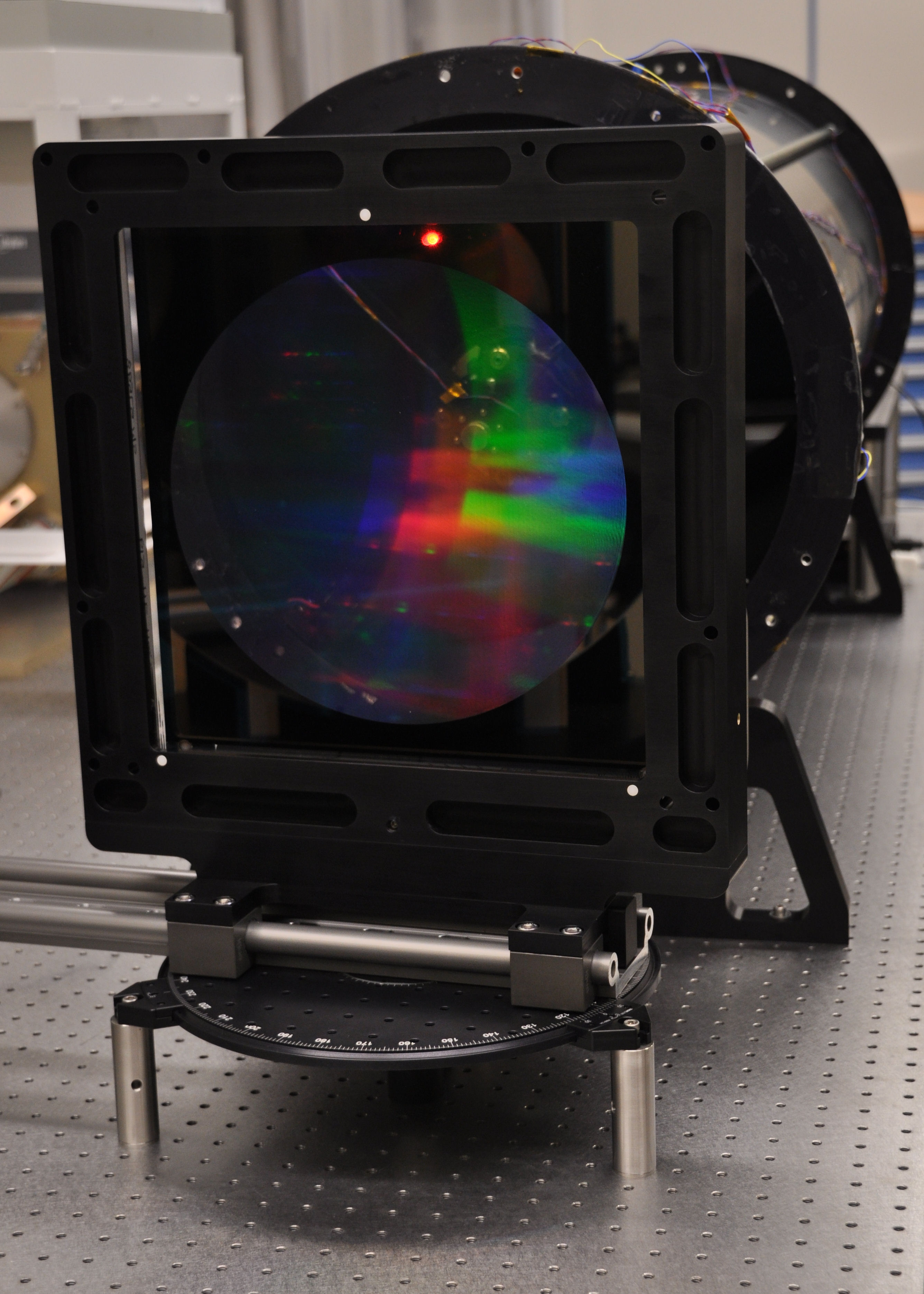}
    \end{tabular}
    \caption{The PFS grating test mount holding one of the prototype gratings.}
    \label{fig:grating_mount}
\end{figure}

Following the collimator in the setup is the grating test mount (see Figure~\ref{fig:grating_mount}).  The grating is held in a frame by a 3-2-1 configuration of delrin pads with opposing spring plungers.  The frame is easily removed from the base for loading and unloading gratings in a horizontal orientation, with handles on the frame and guide pins on the base to make installation safe and repeatable.  A spring pin in the base engages a tab in the frame to temporarily secure the frame to the base until screws can be installed to positively fasten the frame to the base.  The base is mounted to a carriage which rides on a dual shaft rail via four linear bushings, providing smooth and easy positioning of the grating in and out of the beam.  Thought was given to automating the motion with a motor-driven lead screw but in the end a manual rail was deemed sufficient given the expected number of actuations for testing the full suite of PFS gratings.  Mechanical stops are located at each end of the rail to prevent any disasters resulting from overshooting the range of travel.  The rail is mounted to a large-diameter manual rotation stage which allows the angle of the grating to be accurately set for the grating under test.  An outrigger post at the end of the rail provides stability when the mount is positioned out of the beam.

 The final piece of the wavefront measurement setup is a high quality \textlambda/20 optical flat to autoreflect the beam.  The flat was originally procured for the same program as the collimator, and is held in a sling mount with precision tip/tilt adjustment screws.  The required position and azimuthal angle of the flat changes by a large amount depending on the diffracted angle of the grating under test, and sliding the flat across the bench on its leveling screws is cumbersome and potentially hazardous to the glass given stiction with the bench and the tendency for pendulum motion of the flat within the sling.  For these reasons, three ball transfer rollers will be installed in the baseplate of the mirror mount and will ride in curved grooves machined into a large aluminum plate on the bench.  The curved grooves are concentric with the rotation axis of the grating mount; therefore, as the flat is moved laterally into position for different diffraction angles, the surface of the flat remains roughly normal to the beam and should require only minor adjustment using the tip/tilt screws.  The grooved plate is visible between the optical flat and the grating mount in Figure~\ref{fig:wfe_setup}, pushed off to the side.  When installed in the setup it will be located underneath the optical flat.

%%-----------------------------------------------------------
\subsection{Point spread function (PSF) and scatter}
\label{sec:psf}

\begin{figure}[]
    \centering
    \includegraphics[width=\textwidth]{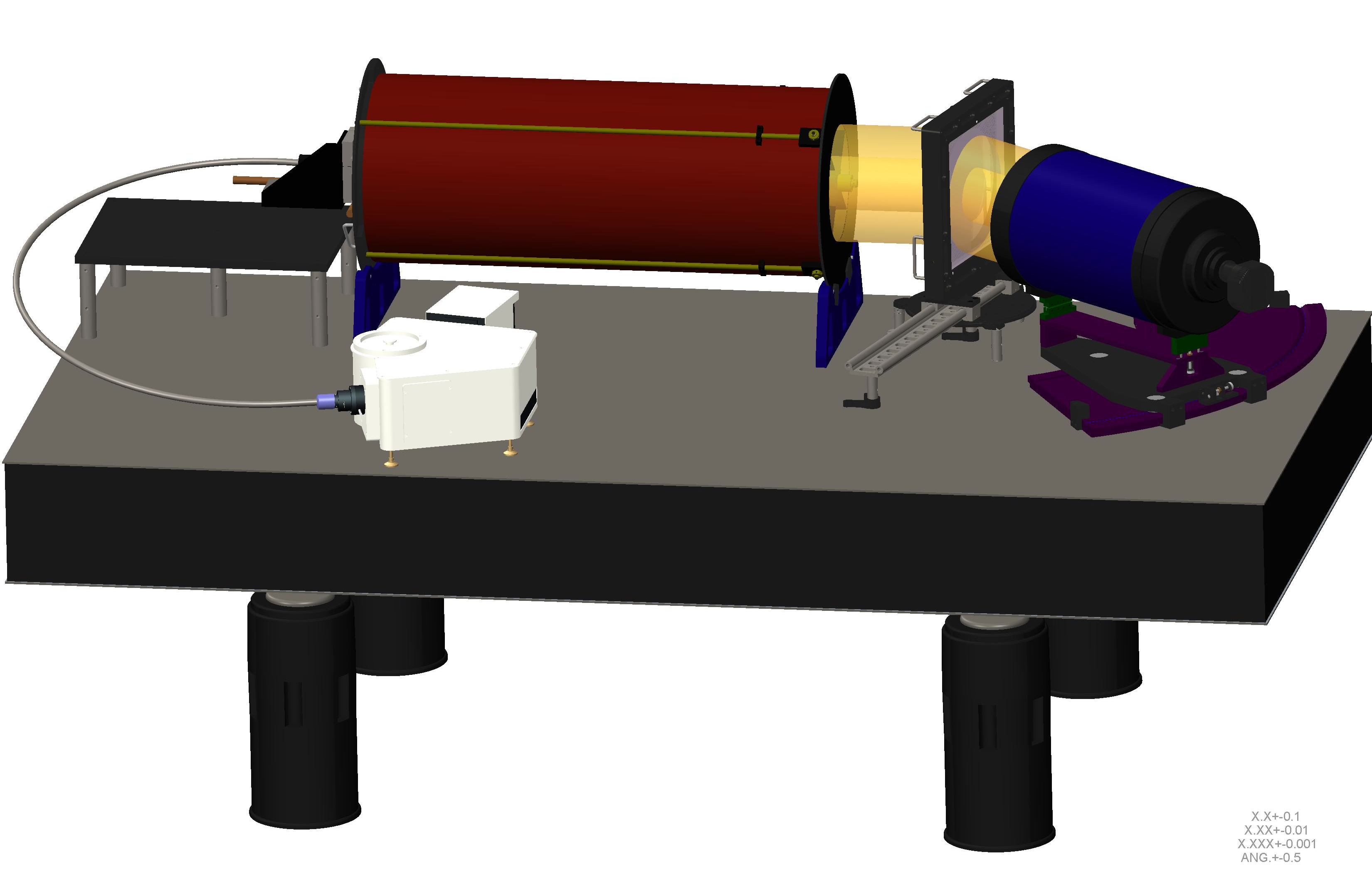}
    \caption{Test setup for PSF and scatter measurements.}
    \label{fig:psf_setup}
\end{figure}

Accurate sky subtraction is necessary for PFS and requires that the spectrograph optical system deliver images with well-controlled energy in the wings of the point spread function (PSF).  The gratings are expected to be the largest single contributor to image spread in each channel, and for this reason we want to have a direct measurement of the effect of each grating on the PSF.  The collimator provides the front end necessary to put a high quality wavefront into the grating under test; we just needed a capable back end to collect the light and focus it without excessive degradation onto an appropriate detector.  The Schmidt-Cassegrain telescopes designed for amateur astronomy offered a cost-effective solution and we procured a 12 inch, f/10 Meade LX200 optical tube assembly (OTA).  A mount has been designed and fabricated which will allow the OTA to ride in the same curved grooves as the optical flat, and also provides precision elevation and azimuthal adjustment once the OTA is roughly positioned in the beam.  A CAD model image of the PSF test setup is shown in Figure~\ref{fig:psf_setup}.

A pinhole at the focal plane of the collimator will be illuminated by a Horiba iHR320 monochromator, via a custom fiber bundle from Fiberguide Industries\footnote{Fiberguide Industries, 1 Bay Street, Stirling, NJ 07980 USA} containing 60 low OH fibers with 800 \textmu m cores, arranged in a 4.0 x 11.1~mm slit pattern at one end and a 7.9~mm diameter circular pattern at the other end.  The monochromator contains a triple grating turret which will cover the full bandpass of PFS from 380~nm out to 1.26~\textmu m.  A thermoelectrically-cooled CCD camera with 6.8~\textmu m pixels from QSI will attach to the OTA.  The CCD of course will not be sensitive at 1.26~\textmu m, but it does have some response longward of 1 \textmu m and we anticipate being able to obtain images out to 1.1 \textmu m.  This should be sufficient for characterizing the PSF of the NIR gratings, but if needed we will procure or borrow a near-IR camera.  The collimator+OTA system provides a magnification of 0.67 and we expect negligible image degradation relative to that of the grating.

In addition to characterizing the grating PSF, we intend to use this setup to measure scattering caused by the gratings.  VPH gratings are known to have low scatter in general, but we would like to be able to make a quantitative measurement for these gratings.  We likely will be dark current limited due to low flux levels far from the core of the PSF, combined with the cooled but not cryogenic CCD, but we do have the very bright laser source of the LUPI which is within the bandpasses of the blue and red gratings.

%%%%%%%%%%%%%%%%%%%%%%%%%%%%%%%%%%%%%%%%%%%%%%%%%%%%%%%%%%%%%
\section{MEASUREMENT RESULTS}

%%-----------------------------------------------------------
\subsection{Diffraction efficiency}

\begin{figure}[h]
    \centering
    \includegraphics[height=9cm]{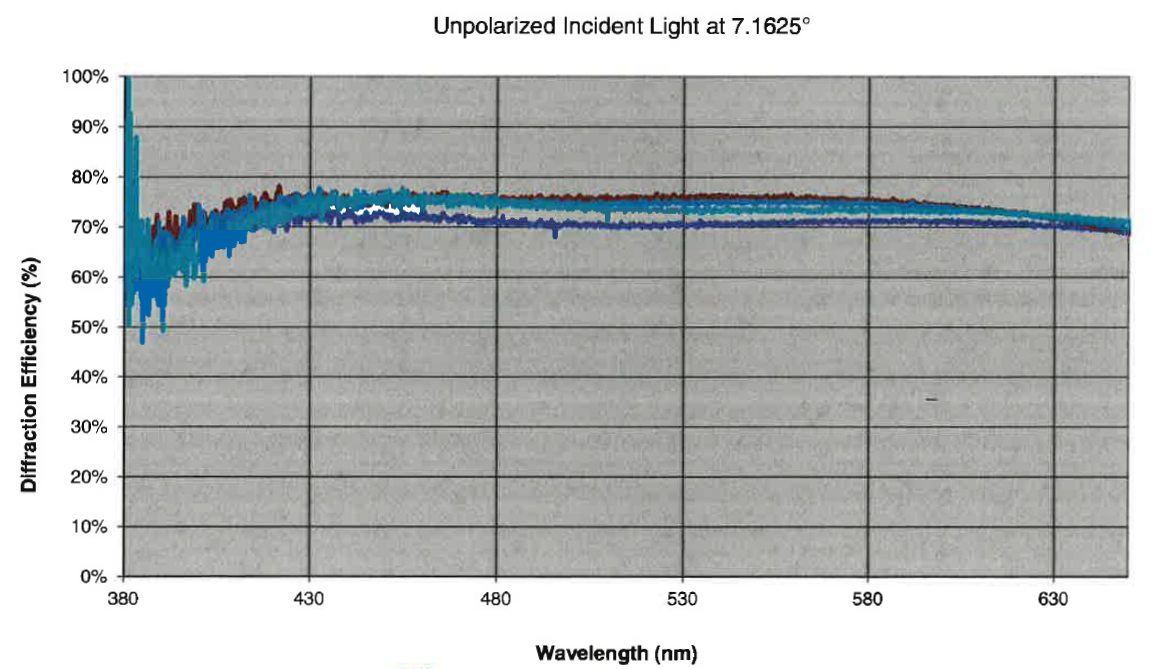}
    \caption{Measured diffraction efficiency of blue grating.}
    \label{fig:blue_diffract}
\end{figure}

\begin{figure}[h]
    \centering
    \includegraphics[height=9cm]{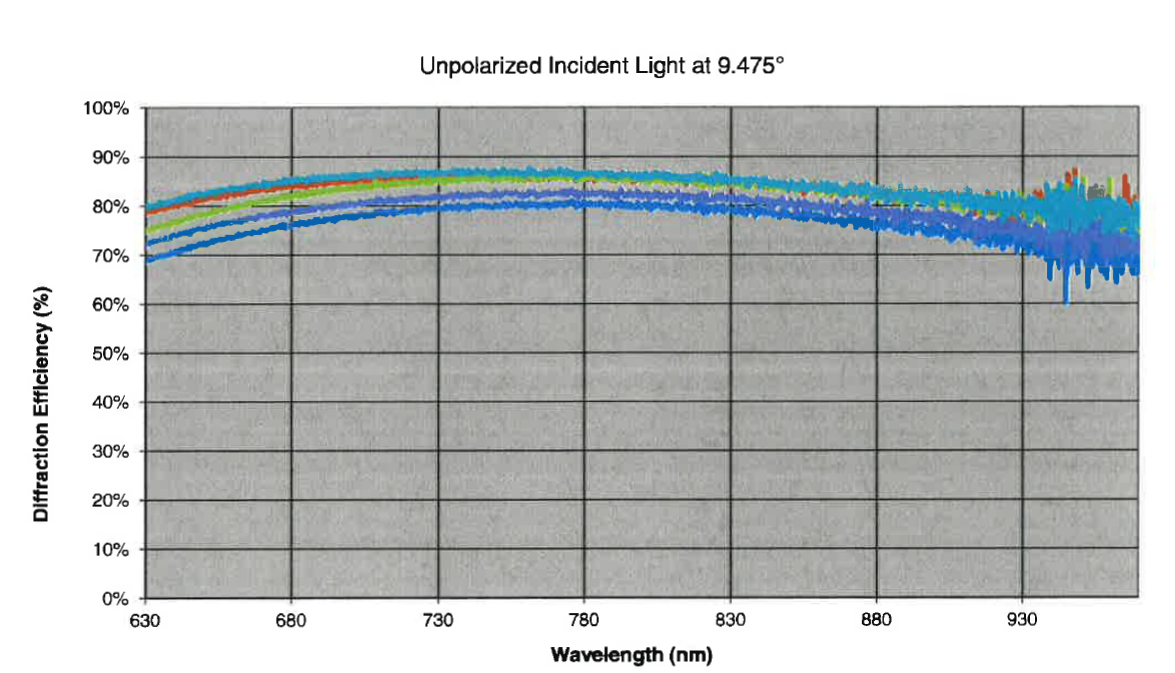}
    \caption{Measured diffraction efficiency of red grating.}
    \label{fig:red_diffract}
\end{figure}

\begin{figure}[h]
    \centering
    \includegraphics[height=9cm]{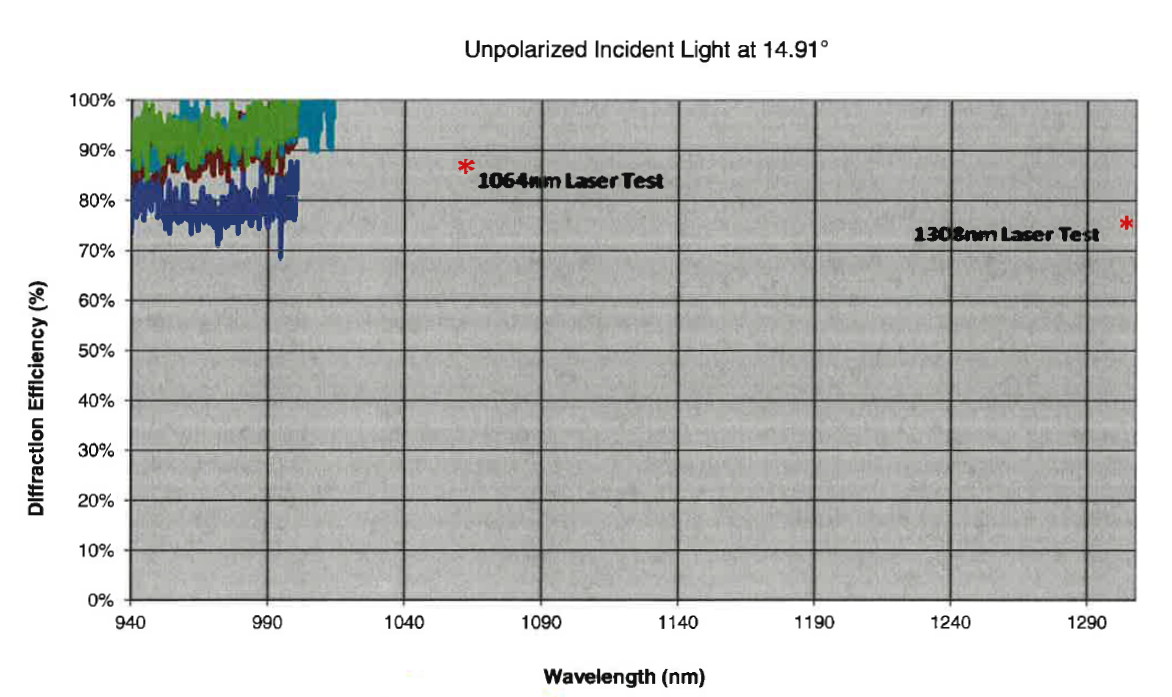}
    \caption{Measured diffraction efficiency of NIR grating.}
    \label{fig:nir_diffract}
\end{figure}

The blue and red prototype gratings were measured at Kaiser using the spectrometer method described in \S \ref{sec:spectrometer}, as the spectral bandwidth of these two gratings are well within the measurement equipment capability. The NIR grating was measured using the laser and ratiometer test setup, described in \S \ref{sec:ratiometer}, using lasers operating at 1064 and 1308 nm. Some data was collected using the spectrometer equipment in the 940 to 925~nm region even though there was significant noise, to give some indication as to performance at the lowest end of the grating operational wavelength range.  Figures~\ref{fig:blue_diffract}, \ref{fig:red_diffract}, and \ref{fig:nir_diffract} show the results of these measurements.  The results are very promising, with the red and NIR gratings having peak efficiencies near 90\%.  The blue grating has a lower peak efficiency but is very flat across the bandpass.  Comparing to the theoretical predictions discussed in \S \ref{sec:parameters}, the blue efficiency appears to match pretty closely the prediction for a 1~degree fringe slant design.  It is expected that the efficiency of the blue grating will be optimized during fabrication of the production gratings.  The red and NIR gratings are within $\sim 10\%$ of the theoretical peak, and closer than that at the ends of the bandpass.  This is very encouraging and has demonstrated the correctness of the design models and their ability to be manufactured.

The system for full aperture efficiency measurements at JHU is nearly complete, and we expect to begin testing the prototype gratings in late summer of this year (2014).

%%-----------------------------------------------------------
\subsection{Wavefront error}

\begin{figure}[h]
    \centering
    \includegraphics[height=9cm]{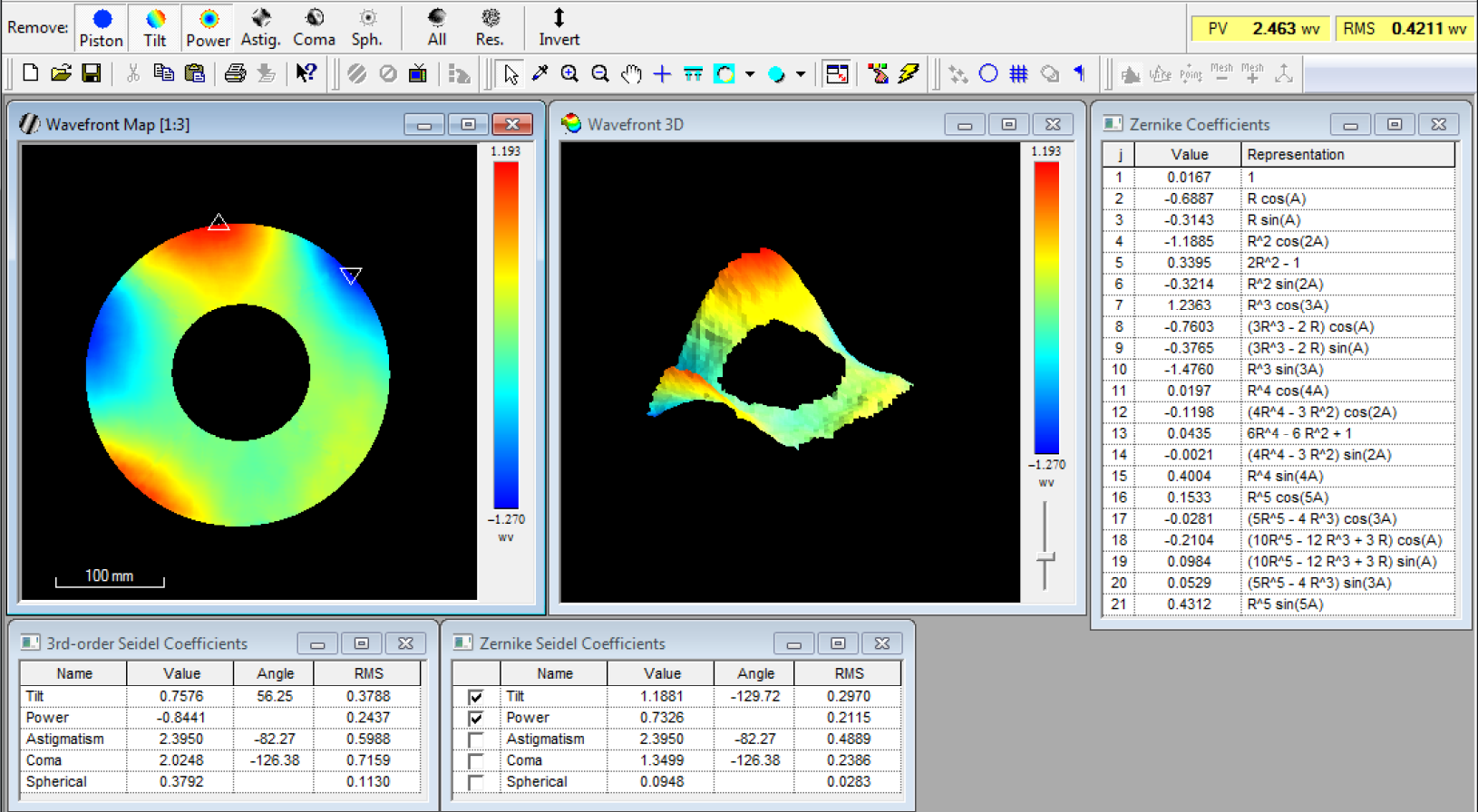}
    \caption{Diffracted wavefront error measured for the blue grating.}
    \medskip
    \label{fig:blue_wfe_jhu}
\end{figure}

\begin{figure}[h]
    \centering
    \includegraphics[height=9cm]{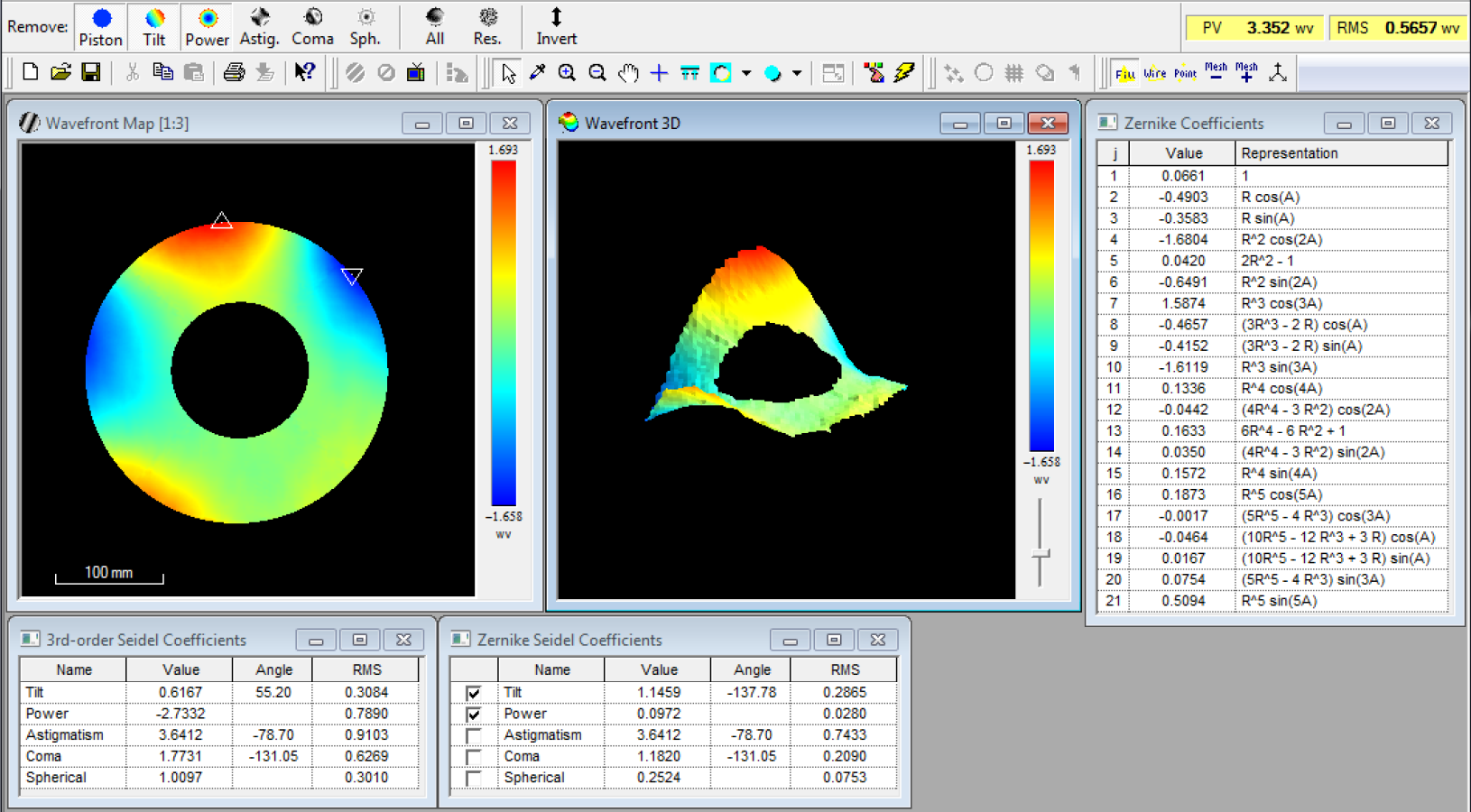}
    \caption{Diffracted wavefront error measured for the red grating.}
    \label{fig:red_wfe_jhu}
\end{figure}

\begin{figure}[h]
    \centering
    \includegraphics[height=9cm]{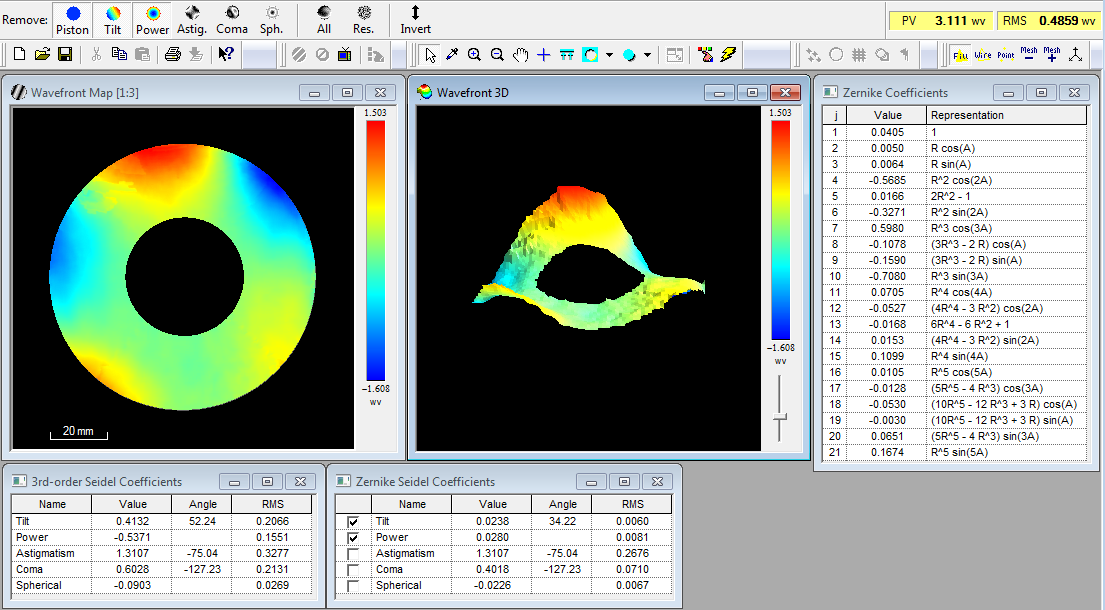}
    \caption{Diffracted wavefront error measured for the NIR grating.}
    \medskip
    \label{fig:nir_wfe_jhu}
\end{figure}

\begin{figure}[h]
    \centering
    \includegraphics[height=7.5cm]{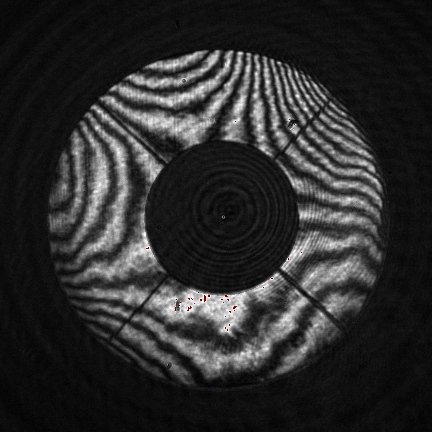}
    \caption{Interferogram of the NIR grating, nulled to show the trefoil evident in the fringe pattern.}
    \label{fig:null_fringes}
\end{figure}

Wavefront error measurements with the Zygo interferometer at Kaiser showed promising results over the central 100~mm of the clear aperture.  The wavefront error for all three gratings was measured to be \textless~0.5 waves P-V at 632.8~nm.  When measured at JHU with the full aperture LUPI setup, we found errors from $\sim$ 2.5 to 3.4 waves P-V (the goal for these gratings was \textless~2 waves P-V over the clear aperture).  Figures~\ref{fig:blue_wfe_jhu}, \ref{fig:red_wfe_jhu}, and \ref{fig:nir_wfe_jhu} show the results from the wavefront measurements of the three gratings.  What is really interesting is the characteristic and very repeatable presence of trefoil distortion in the diffracted wavefront of all three gratings.  The clocking of the distortion is identical across the set, and this distortion is by far the largest error in the wavefront.  Figure~\ref{fig:null_fringes} shows an interferogram at null for the NIR grating, and the trefoil distortion is very evident in the fringe pattern.

\begin{figure}[h]
    \centering
    \includegraphics[height=9cm]{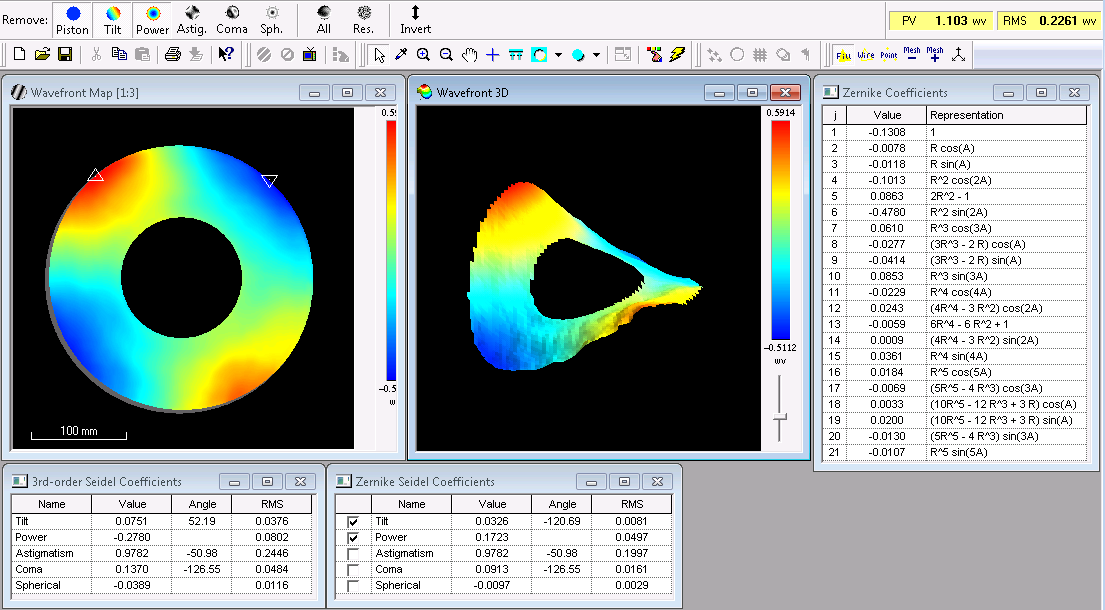}
    \caption{Transmitted wavefront error measured in 0\textsuperscript{th} order for the NIR grating.}
    \label{fig:nir_wfe_zero_jhu}
\end{figure}

In an effort to track down the source of the distortion, the transmitted wavefront of the NIR grating in zeroth order was measured with the identical setup.  Figure~\ref{fig:nir_wfe_zero_jhu} shows the results from this measurement, which indicated the presence of about 1 wave of pure astigmatism and not much else.  This is the type of error, and at about the level, that might expected to arise during the bonding of the cover plate to the grating substrate.  Further investigation with the LUPI revealed that the front and rear surfaces of the grating assembly appear to be reasonably flat with no hint of the trefoil distortion.  Fringes were also obtained off the chrome mask surface and showed no sign of trefoil or other significant distortion.

Because the trefoil only appears in the diffracted wavefront and is so repeatable from grating to grating, our strong suspicion was that the wavefront error was either in the exposure optics or from distortions in the mount used to hold the grating during exposure.  Recent investigations at Kaiser have revealed the presence of misalignment and possibly some pinching in the exposure setup, and efforts are underway to isolate and resolve the source of error.

%%%%%%%%%%%%%%%%%%%%%%%%%%%%%%%%%%%%%%%%%%%%%%%%%%%%%%%%%%%%%
\section{SUMMARY}

Three prototype VPH gratings for PFS have been fabricated and tested, and we have presented the results here.  The diffraction efficiency of the gratings appears to be quite good, and these gratings were not completely optimized during processing.  Thus, we expect that the production gratings will be a bit better, especially for the blue channel.  Wavefront testing revealed the unexpected presence of a trefoil distortion at the level of a few waves P-V.  The source of this error is suspected to be in the alignment of the holographic exposure optics, and this issue should be resolved in the near future, before starting the production gratings.  The chrome pupil masks were successful for the prototype gratings but continue to be problematic in terms of obtaining good quality chrome coatings, causing long lead times for the cover plates needed before grating fabrication can begin.  The prototype gratings have also been very useful in debugging our test setups and procedures.

%%%%%%%%%%%%%%%%%%%%%%%%%%%%%%%%%%%%%%%%%%%%%%%%%%%%%%%%%%%%%
\acknowledgments     %>>>> equivalent to \section*{ACKNOWLEDGMENTS}

We gratefully acknowledge support from the Funding Program for World-Leading Innovative R\&D in Science and Technology (FIRST), program: ``Subaru Measurements of Images and Redshifts (SuMIRe),'' CSTP, Japan

%%%%%%%%%%%%%%%%%%%%%%%%%%%%%%%%%%%%%%%%%%%%%%%%%%%%%%%%%%%%%
%%%%% References %%%%%

\bibliography{report}   %>>>> bibliography data in report.bib
\bibliographystyle{spiebib}   %>>>> makes bibtex use spiebib.bst

\end{document}